\documentstyle[11pt,newpasp,twoside]{article}
\def\edcomment#1{\iffalse\marginpar{\raggedright\sl#1\/}\else\relax\fi}
\reversemarginpar

\begin{document}

\title{Introductory Remarks on The Extragalactic Infrared
Background and its Cosmological Implications}
\author{P. J. E. Peebles}
\affil{Joseph Henry Laboratories, Princeton University,
Princeton, NJ 08544}

%\authoremail{pjep@pupgg.princeton.edu}
 
\begin{abstract}
I review the assumptions and observations that motivate the
concept of the extragalactic cosmic background radiation, and the
issues of energy accounts and star formation history as a
function of galaxy morphological type that figure in the
interpretation of the measurements of the extragalactic infrared 
background. 
\end{abstract}

\section{Fundamental Assumptions}

It is useful to begin by recalling the basic assumptions and 
observations that lead us to the concept of the
extragalactic cosmic background radiation, as opposed to
radiation surface brightness that may be some highly variable
function of position and direction. 

Deep counts of objects detected at a broad range of wavelengths,
from gamma ray sources to radio galaxies, are close to isotropic  
across the sky. It is an excellent bet 
therefore that the integrated radiation from observed sources
plus those too faint to be detectable as individual objects also
is quite to isotropic. This allows us to think of the local
extragalactic radiation background as a function of one variable,
the radiation energy density $u_\nu$ per interval of frequency
$\nu$. The fluctuations around the mean as a function of position
in the sky are important too, as a measure of large-scale
structure, but $u_\nu$ is the center of attention in these
Proceedings.  

The argument for large-scale homogeneity --- against a
universe with a radial density gradient and us at the
center --- is less direct but I think persuasive; my review of
the considerations is in Peebles (1993). If we are persuaded then
we conclude that within our Hubble length space is filled with a
near uniform sea of radiation with spectral energy density
$u_\nu$: the cosmic extragalactic background radiation.

If the propagation of the radiation is described by a metric
theory then it satisfies the Liouville or brightness theorem.
If the metric describes a homogeneous isotropic spacetime then
the geometry is fixed by the expansion factor $a(t)$, a
function of the proper world time $t$ alone, together with the
radius of curvature $a(t)R$ of sections of constant time, where
the comoving radius $R$ is a constant. In this spacetime the
radiation energy density $u(t)=\int d\nu\, u_\nu$ integrated over
frequency at time $t$ is an integral over the history of
production and absorption of radiation,   
\begin{equation}
u(t) = \int ^t dt'\, j(t')\, (a(t')/a(t))^4.
\end{equation}
At tine $t'$ the net rate of production of radiation (emitted
minus absorbed) 
per unit proper volume is $j(t')$, and $j(t')\, (a(t')/a(t))^3$
is the rate of production of energy per comoving volume
normalized to the time $t$ of observation. The remaining factor
in the integrand, $a(t')/a(t)=(1+z)^{-1}$, where $z$ is the
redshift at the epoch $t'$ observed at time $t$, represents
energy lost due to the cosmological redshift.  

If spacetime were static, $a$ independent of time, 
equation~(1) says $j$ could not have been constant: 
there would have to have been a characteristic time at
which star formation commenced. The point, associated with
the name Olbers, is not often mentioned now; an edifying
discussion is to be found in Bondi (1960). In  
the classical steady state cosmology (which also is well described
by Bondi) the universe is expanding, $a\propto e^{Ht}$, where
$H$ is Hubble's constant. This makes the integral converge even
when $j$ is constant, stars forming at a fixed mean rate per
physical volume back to the indefinitely remote past. But we know
now this is not a viable picture: Cowie and Lilly describe in
these Proceedings observations of galaxies and an intergalactic
medium at high redshift that are distinctly different from what
is observed nearby; the more youthful appearance of objects 
at high redshift agrees with the interpretation that they are
seen closer to the time when the structure we see started
forming. In the general relativistic 
Friedmann-Lema\^\i tre model with a classical stress-energy
tensor that satisfies $\rho +3p>0$ the integral in
equation~(1) has to have a lower 
limit, at the singular start of expansion at $a(t)=0$. In the
eternal inflation scenario (Linde~1990) this unsatisfactory
situation is relieved by the return to a steady state philosophy:
the lower limit to the integral extends back along our world line
to the remote past. 

\section{Cosmic Energy Densities}

Let us consider now the interpretation of the radiation
background under the standard relativistic cosmology. 
Evolution after inflation --- or whatever produced the
initial conditions for the present state of our expanding
universe --- was accompanied by exchanges of energy among
different forms. An accounting of the integrated results of the 
transactions at the present epoch offers a measure
of cosmic evolution, and in particular it informs our
interpretation of the infrared background.   

The estimates in Table~1 are expressed in units of the
Einstein-de~Sitter value, $\rho _c = 3H_o^2/(8\pi G)$, at Hubble
constant $H_o=70\pm 10$ km~s$^{-1}$~Mpc$^{-1}$. That is, these numbers
are contributions to the cosmological density parameter. The
first set of numbers, labeled primeval, are thought to have been
fixed by physical processes operating in the early universe, well
before stars and galaxies started forming; the second set are
estimates of the effects of the formation and evolution of
structure on scales ranging from clusters of galaxies down to
star remnants. 

The accounting in this table accepts the evidence for a
Friedmann-Lema\^\i tre model that is close to cosmologically
flat, the stress-energy tensor being dominated by a term that
acts like Einstein's cosmological constant, $\Lambda$. The
next most important term appears to be some form of nonbaryonic
dark matter. The baryon density in the third line agrees with the 
theory of the origin of the light elements in the early universe,
with the fluctuation spectrum of the 3~K thermal background
radiation --- within reasonable-looking uncertainties (eg. Hu et
al. 2000) --- and with the observational constraints on the
baryon budget (Fukugita, Hogan, \&\ Peebles 1998). The baryon
entry seems secure to 
30\%\ or so, a truly remarkable advance. It is a measure of the
state of our subject that the two largest entries are
conjectural. The evidence for low pressure dark matter at about
the density indicated in the 
table is compelling if we accept general relativity theory (and
hence the inverse square law for gravity); the evidence for
$\Lambda$ or its near operational equivalent is strong if we
accept the adiabatic CDM model for structure formation. Work in
progress promises to establish more tests of general relativity
theory applied on the scale of the Hubble length, of the
cosmology, and of the theory of structure formation. The 
results certainly will be searched for potential insights into
the enigmatic leading entries in the table.

\begin{table}
\begin{center}
\begin{tabular}{lc}
\multicolumn{2}{c}{Table 1.\qquad The Cosmic Energy Account}
\vspace{1.5mm} \\
\tableline
\vspace{-3.5mm} \\
\tableline
\vspace{-3mm} \\
 & $\Omega$ \\
\vspace{-3mm} \\
\tableline
\vspace{-3mm} \\
\multicolumn{2}{c}{Primeval} \\
\vspace{-3.5mm} \\
\tableline
\vspace{-3mm} \\
$\Lambda$/Quintessence/dark energy \qquad\qquad & 
$10^{-0.1\pm 0.1}$ \\
\vspace{-3.5mm} \\
low pressure nonbaryonic matter & $10^{-0.75\pm 0.25}$ \\
baryons & $10^{-1.3\pm 0.1}$ \\
relict neutrinos & $10^{-2.4\pm 0.8}$ \\
thermal radiation & $10^{-4.15}$ \\
gravitational binding energy & $\sim -10^{-6}$ \\
\vspace{-3mm} \\
\tableline
\vspace{-3mm} \\
\multicolumn{2}{c}{Products of Structure Formation} \\
\vspace{-3.5mm} \\
\tableline
\vspace{-3mm} \\
gravitational binding energy: \\
\qquad relativistic & $\sim - 10^{-5.4}\epsilon $ \\
\qquad stars & $\sim - 10^{-7.8}$ \\
\qquad galaxies & $\sim - 10^{-8.3}$ \\
nuclear binding energy:\\
\qquad helium & $-10^{-5.6\pm 0.5}$ \\
\qquad heavy elements & $-10^{-5.9\pm 0.3}$ \\
X-gamma radiation & $\sim 10^{-8.5}$ \\
optical/near ir radiation & $\sim 10^{-6}$ \\
far ir/sub mm radiation & $\sim 10^{-6}$ \\
\vspace{-3mm} \\
\tableline
\end{tabular}
\end{center}
\end{table}

The entry for relict neutrinos that broke thermal
equilibrium with the background radiation at $z\sim 10^{10}$
assumes the tau neutrino mass is no smaller than 0.03~eV, from
atmospheric neutrinos (Super-Kamiokande Collaboration 2000),
and no larger than about 1~eV, to avoid undue effect on structure
formation (Klypin et al. 1993). The accepted provenance of the
3~K thermal radiation and its related neutrinos ---
entropy originating near the end of inflation --- is conjectural. 
The observed peak in the angular fluctuation spectrum of this 
radiation, at the length scale set by decoupling at $z\sim 1000$,
is good evidence this radiation is primeval, present well 
before the observed stars and galaxies could have started forming.

The meaning of the last of the primeval entries is illustrated by
the comparison of two Friedmann-Lema\^\i tre models, both containing 
only cold dark matter particles with the same particle mass, and
with the same comoving radius of curvature measured relative to
the mean distance between particles. In one model the mass
distribution is close to homogeneous; in the other model 
primeval curvature fluctuations have placed most of 
the dark matter in gravitationally bound nonrelativistic halos.
At a given mean particle number density the mean
mass density is smaller in the latter model by the amount of
the mean halo gravitational binding energy: the sum of the
kinetic energy of proper motion of the particles relative to the
general expansion and the gravitational potential energy relative 
to a homogeneous mass distribution. Thus when the particle number
densities are the same the physical radii of curvature are the 
same in the two models, and the expansion rate, from the
relativistic expression 
$(\dot a/a)^2 = 8\pi G\rho /3 \pm 1/(aR)^2$, is lower in the model
with halos. This gravitational binding energy in the model with
halos is primeval in the sense that it is a consequence of the
gravitational growth of structure out of given initial
conditions; there is no energy transaction when the halos
collapse to form bound systems. We have good evidence that this 
gravitational growth of structure is responsible for the
large-scale structure of our universe; most dramatic is the
consistency of the angular fluctuation spectrum of the 3~K
background with the simple 
adiabatic CDM model for the gravitational instability picture 
(eg. Hu et al. 2000 and references therein). In this CDM model
radiation pressure suppresses the growth of density fluctuations
on scales less than the matter-radiation Jeans length at
decoupling, $\sim 10$~Mpc, lowering the binding energy, and the 
dissipation of the small-scale pressure waves adds a little
entropy. The number in the table for the primeval gravitational
binding energy assumes the rms peculiar velocity is comparable
to that of the Local Group relative to the 3~K background, 
$~\sim 600$ km~s$^{-1}$, in matter at the sum of the mass
densities in the second and third entries.  

Many have commented on the small differences of values of
successive entries in this first group, which may in part
reflect the anthropic consideration that these parameters can be
adjusted so we could not have been here to measure them, may
in part result from physical relations to be discovered, and
may in part be pure coincidence. Similar remarks apply to the
second group of entries, of course. 

The first entry in this second group is based on the contribution
to the density parameter by the mass bound in compact nuclei of
galaxies (from the review by Fabian 2000). If these objects are
relativistic black holes, mass may flow into them 
without the emission of energy in radiation or jets. For our
purpose this accretion without emission represents an exchange of
small particles for large ones, both gravitationally bound to
the massive halos of galaxies, and it has no effect on the
accounting in Table~1. The same applies to the kinetic energy of
a jet that is dissipated within the galaxy. The factor
$\epsilon$ in the table is the mean fraction of energy in
radiation and jets that has left the gravitationally bound
halos within which these compact objects live. 

Most of the baryons are thought to be in intragroup and
intracluster plasma, distributed in about the same way as the
dark matter. The gravitational binding energy per unit mass
consequently is about the same, and, as we have noted, in the
adiabatic CDM model this primeval binding energy appeared without
transfer of energy to some other form. The dominant energy
exchange in this component of the baryons may be the thermal
bremsstrahlung and emission line X-ray 
radiation from intracluster plasma; less important transactions
include the energy exchanges due to galactic winds and the energy
exchanges between baryons and dark matter via fluctuations in the 
gravitational potential.   

The formation of baryon-rich spheroids and disks of galaxies
requires about the same dissipative production of binding energy 
per unit mass as does star formation. Here too one can think of
subdominant energy transactions: magnetic fields produced by
galaxy dynamos and then blown out with the galactic winds, and
escaping cosmic rays. The largest transaction likely was the 
contribution to the cosmic background radiation.  

The entry for the binding energy in atomic nuclei heavier than
helium is from Fukugita (2000); it uses
the baryon budget of Fukugita, Hogan, \&\ Peebles (1998)
at $H_o = 70$ km~s$^{-1}$~Mpc$^{-1}$, and assumes the heavy
element abundance is $\sim 0.4$ times Solar in the intracluster
plasma and $\sim 0.1$ times Solar in the intragroup plasma. The
production of helium in stars assumes the ratio of production
of helium and heavier elements is $1\la\Delta Y/\Delta Z\la 4$.

The energies in the optical/near ir and far~ir/sub-mm
extragalactic radiation backgrounds are taken from Pei, Fall, \&\
Hauser (1999). These measurements, and their interpretations, are
the subject of these Proceedings; a few comments suggested
by the entries Table~1 will be noted here.

The optical extragalactic background at $\lambda\sim $5000~\AA\
likely is dominated by starlight from low redshift, because the
spectra of most galaxies decrease toward the ultraviolet. If so 
this background is not sensitive to the 
parameters of the cosmological model, but it is of 
considerable importance to cosmology as a test for stars in the
extreme outer halos of galaxies or in objects with effective
radii or surface brightnesses too small to be included in galaxy
counts (Arp 1965; Peebles 1971). The measurement of the
extragalactic contribution to the light of the night sky 
has a long history (eg. Dube, Wickes, \&\ Wilkinson 1979;
Bernstein 2000). It shows we have not grossly underestimated the
density of starlight, but further advances in the measurements
will be followed with interest. 

The estimate of the energy released by star formation is less
than the upper bound on the background radiation between the
optical/near ir and far~ir/sub-mm peaks, not an observationally
promising situation. 

The energy released during dissipative
contraction to the central baryon-dominated parts of the
galaxies could make an interesting contribution to the X-ray
background. The contribution by spheroids depends on the mode of
formation, whether by dissipative contraction of plasma followed
by star formation or by near dissipationless
merging of dense star clusters. Wyse (1999) reviews the
observational considerations on which might be closer to what
happened. If the latter, and if the star cluster building blocks
had small individual escape velocities, the radiation
accompanying the dissipative assembly of the star clusters could
be lost in the uv/optical background. Disks likely formed by
dissipative settling, and might be a significant source for the
soft X-ray background.   

The energy density in the X-ray background is three orders of
magnitude down from the estimate of the energy density in 
black holes in galactic nuclei. Fabian (2000), Hasinger (2000)
and others note that even with the $1+z$ redshift dimming
(eq.~[1]), and a low radiation efficiency, $\epsilon$, there is an
ample budget for significant contributions by AGNs to the optical
and sub-mm backgrounds. 

About half the extragalactic B-band background
radiation comes from the disks of spiral galaxies. The dust in
disks absorbs a substantial fraction of the starlight, so the
contribution to the energy density in the sub-mm radiation
background from starlight absorbed and reradiated by dust
would be expected to be comparable to the contribution to 
the optical/near infrared radiation by unabsorbed starlight. This
is in line with the background measurements discussed by Hauser
and others at this meeting. The entry for the binding energy in
atomic nuclei 
is comparable to the measured energy in the infrared background. 
And with Pettini's (1999) calibration the time integral of the
observed star formation history agrees with the present mass
density in stars. These three results could be taken to suggest
we are getting close to a reconciliation of our accounts.
There are some complications to consider, however, as discussed
next.    

\section{Star Formation Histories}

Michael Fall and Piero Madau discuss in these Proceedings
an important observational and conceptual advance, the
establishment of a global star formation history. This history
can be compared to the record of the evolution of heavy element
abundances, and to the resulting extragalactic infrared 
background. I will present reasons for thinking complications in
the history may require a generalization of the star formation
history to a function of two variables, world time and
environment. The latter might have just three values: normal
$L\sim L_\ast$ galaxy bulges, galaxy disks, and everywhere
else. Even with this simple second parameter the assembly of
useful observational constraints would be a messy project,
but it may be necessary for concordance in the
interpretation of a more sophisticated version of Table~1.  

To begin, suppose we choose cosmological parameters (and the
range of values now under popular discussion is not that broad)
so as to fix the world time $t$ as a function of   
redshift. Then the Madau plot (Pei \&\ Fall 1995; Madau et al.
1996) --- the observed star formation rate $dM_\ast /dt$ per
unit comoving volume as a function of redshift --- could be
replotted as the product $t\, dM_\ast /dt$ as a function of 
$\log t$ (or, what is equivalent,
$z\, (dM_\ast /dt)\, (dt/dz)$ against $\log z$). The   
motivation is the same as for the representation of the energy 
density of the extragalactic background as 
$\nu u_\nu =\lambda u_\lambda$: in a 
semi-logarithmic plot the area under the curve is the
contribution to the integral over the independent variable per
logarithmic interval of the variable. That is, this way to
represent the star formation history gives us a picture of when 
there were significant contributions to the net mass in stars.
One would see that about half the observed star formation was
relatively recent, at redshift $z<1$. Where are these young
stars? The answer may well be complicated; I will mention two
simple possibilities.  

First, the star formation observed at $z<1$ might be concentrated
somewhere other than in the normal textbook galaxies. Maybe this has
something to do with the enigmatic fading faint blue galaxy
population; we don't know where they ended up either. In this
scenario most stars may have formed at $z<1$ and produced most of
the metals and most of the extragalactic infrared radiation. But 
by assumption the extragalactic infrared background would have no
direct relation to the origin and evolution of the populations of
normal galaxies.  

A second scenario is that most of the stars that formed at $z<1$ 
end up in normal $L\sim L_\ast$ galaxies. Then the evidence as I
understand it is that these stars would have to be subdominant
additions to the predominantly older star populations. We might
recall that this 
evidence includes these four points. a. About two thirds of
the star mass in the high surface density $L\ga L_\ast$ galaxies
is in spheroids. The evidence I have been hearing is that most of
these spheroid star populations are a good deal older than 
$z\sim 1$. b. Galaxy counts as a function of redshift at $z\la 1$
are consistent with near passive evolution of the star populations
already present at 
$z=1$. c. It seems to be agreed that the present population of
normal $L\ga L_\ast$ galaxies by and large were in place, with
near familiar morphologies, and close to the present-day comoving
number densities, at $z=1$. d. The normal-looking $L\sim L_\ast$
spirals and ellipticals observed at $z\sim 1$ are close to the 
Tully-Fisher and fundamental plane dynamical relations, after
correction for passive evolution, consistent with the idea that
these are full-grown textbook galaxies. I think we
must conclude that in this scenario the observed star 
formation history does not include the main event: the bulk of
the star formation would have to be off scale to larger redshift,
or maybe present at the redshift range of the Madau plot but
heavily obscured by dust and in concentrations small enough to
have avoided too many SCUBA detections. 

We have a constraint on this hypothetical ``main event'' from the
production of heavy elements. Pei, Fall, \&\ Hauser (1999)
estimate that the observed star formation rate integrated
from high redshift to $z\sim 2.5$ is about enough to account for
the observed accumulation of mass in heavy elements at this
redshift, while Pettini (1999) and Pagel (1999) argue that the
observed mass in heavy elements at $z\sim 2.5$ may actually be 
significantly less than what might have been expected from the
seen amount of star formation up to this epoch. Under the former
estimate the 
``main event'' requires a special conjecture: we have to suppose
the heavy elements associated with the substantial early star
formation are hidden, perhaps in dense clouds.
Under the latter estimate heavy elements have to be 
hidden at $z\sim 2.5$, so it would not be such a great stretch to
imagine the heavy elements associated with the ``mail event'' are
hidden too. Advances in the subtle analysis of the constraints on
the star formation history, including the relation between star
formation and heavy element production, will be fascinating to
follow.

\section{Concluding Remarks}

It has been widely and properly noted that we have 
significant observational evidence galaxies as we know them
formed recently, at $z\la 1$: the energy in the infrared
extragalactic background, without large redshift dimming, is
comparable to the binding energy of the heavy elements in normal
$L\ga L_\ast$ galaxies, and the time integral of the seen 
star formation rate, which implies most stars are young, agrees
with the seen mass in stars. But I have argued other
lines of evidence indicate the cohort of 
stars and their heavy elements that formed at $z<1$ had to have
ended up somewhere other than the high surface brightness parts
of normal galaxies, or else are subdominant additions to the
normal galaxies. True coincidences happen, of course: we have a 
sensible case for similar contributions to the sub-mm background
by two quite different sources --- the radiation from stars and from 
AGNs --- with quite different fractions processed through dust. 

The lesson is that untangling the relations among the  
extragalactic radiation background, the histories of star formation
in different environments, and the accumulations of stars and
heavy  elements, is likely to be a complicated operation. This is
hardly surprising, considering that we live in a
complicated universe. But the suite of observational evidence in
hand is rich, rapidly growing, and fascinating to see sorted out. 

\acknowledgements

I am grateful to Masataka Fukugita for his help in constructing
Table~1, to David Hogg for his help with the astronomy, and to
Michael Fall and Piero Madau for their helpful comments on the
paper. This work was supported in part by the US National Science
Foundation.


\begin{references}

\reference{} Arp, H. C. 1965, ApJ, 142, 402

\reference{} Bernstein, R. A. 2000, to be published

\reference{} Bondi, H. 1960, Cosmology (Cambridge
University Press), 2$^{\rm nd}$ edition

\reference{} Dube, R. R., Wickes, W. C., \&\  Wilkinson, D. T. 
1979, ApJ, 232, 333.

\reference{} Fabian, A. C. 2000, astro-ph/0001178

\reference{} Fukugita, M. 2000, private communication

\reference{} Fukugita, M., Hogan, C. J., \&\ Peebles, P. J. E.
1998, ApJ, 503, 518

\reference{} Hasinger, G. 2000, astro-ph/0001360

\reference{} Hu, W., Fukugita, M., Zaldarriaga, M., \&\ Tegmark,
M. 2000, astro-ph/0006436

\reference{} Klypin, A., Holtzman, J., Primack, J., \&\ Reg\"os,
E. 1993, ApJ, 416, 1

\reference{} Linde, A. D. 1990, Particle Physics and Inflationary
Cosmology (Harwood Academic)

\reference{} Madau, P, Ferguson, H. C., Dickinson, M. E.,
Giavalisco, M., Steidel, C. C., \&\ Fruchter, A. 1996,
MNRAS, 283, 1388

\reference{} Pagel,  B. E. J. 1999, in Galaxies and the Young
Universe II, ed. H. Hippelein, astro-ph/9911204

\reference{} Peebles, P. J. E. 1971, Physical Cosmology
(Princeton University Press), p. 59

\reference{} Peebles, P. J. E. 1993, Principles of Physical
Cosmology (Princeton University Press)

\reference{} Pei, Y. C. \&\ Fall. S. M. 1995, ApJ, 454, 69

\reference{}  Pei, Y. C., Fall, S. M., \&\ Hauser, M. G.
1999, ApJ, 522, 604

\reference{} Pettini, M. 1999, in Chemical Evolution from Zero to
High Redshift, eds. J. Walsh \&\ M. Rosa, astro-ph/9902173

\reference{} Super-Kamiokande Collaboration 2000, astro-ph/0009001

\reference{} Wyse, R. F. G. 1999, in The Formation of Galactic
Bulges, eds. C. M. Carollo, H. C. Ferguson, \&\ R. F. G. Wyse
(Cambridge University Press)

\end{references}
\end{document}